\documentclass{ws-ijmpcs}

\begin{document}

\markboth{Kepler et al.}
{White Dwarf Stars}

%
\catchline{}{}{}{}{}
%

\title{White Dwarf Stars}
\author{S. O. Kepler,
Alejandra Daniela Romero, Ingrid Pelisoli \& Gustavo Ourique}

\address{Departamento de Astronomia, Instituto de F\'{\i}sica,\\
Universidade Federal do Rio Grande do Sul,\\
91501-970 Porto Alegre, RS, Brazil\\
kepler@if.ufrgs.br}


\maketitle

\begin{history}
\received{3 Feb 2017}
\end{history}

\begin{abstract}
White dwarf stars are the final stage of most stars, born single or in multiple systems. 
We discuss the identification, magnetic fields, and mass distribution for white dwarfs detected from spectra obtained by the Sloan Digital Sky Survey up to Data Release 13 in 2016, 
which lead to the increase in the number of spectroscopically identified white dwarf stars from 5\,000 to 39\,000.
This number includes only white dwarf stars with $\log g\geq 6.5$ stars, i.e., excluding the Extremely Low Mass white dwarfs, which are necessarily the
byproduct of stellar interaction.
\end{abstract}

\ccode{PACS numbers:}

\section{Introduction}

White dwarf stars are the final evolutionary state of stars with initial masses up to 8.5--10.6~M$_\odot$\cite{Woosley15}, corresponding to 95 -- 97 \% of all stars.
The fraction
depends on the stellar metallicity, which affects both the Initial-Mass-Function and the Initial-to-Final-Mass Relation. 
For single stars, the minimum mass of a present day white dwarf is around 0.30--0.45~M$_\odot${}\cite{Kilic07}, 
because progenitors that would become lower mass white dwarfs have main sequence evolution time larger than the age of the Universe. 
Such masses correspond, considering the mass-radius relation of white dwarfs, to a minimal $\log g\simeq 6.5$. 
Evolutionary models e.g. by Ref.~\refcite{Romero15} indicate that the maximum surface gravity for main sequence A stars, 
which have similar optical spectra to DA white dwarfs, corresponds to  $\log g \leq 4.75$, even for very low metallicity.
There is therefore a gap between low mass white dwarfs and main sequence stars, $4.75~\leq~\log~g~\leq~6.5$.

Most white dwarfs do not generate energy from nuclear fusion, but radiate due to residual gravitational
contraction. Because of the degenerate equation of state, contraction is accompanied by a loss of thermal energy instead of increase
as in the case of ideal gases; the evolution of white dwarfs is therefore often simply described as cooling. The radius of an average white
dwarf star is of the same order of the Earth's radius, which implies that they have small surface area, resulting in very large
cooling times; it takes approximately $10^{10}$ years for the effective temperature of a normal mass white dwarf to decrease
from $100\,000$~K to near $5\,000$~K. Consequently,
the cool normal mass ones are still visible and among the oldest objects in the Galaxy\cite{GarciaBerro16}. Therefore, studying white dwarfs is extremely important to
comprehend the processes of stellar formation and evolution in the Milky Way\cite{Winget87,Campos16}. 

The progenitors of white dwarfs lose most of their envelope in the giant phases,
where mass loss depends on metallicity.
If the remainder H mass were above $\simeq 10^{-4} M*$, or the He mass above $\simeq 10^{-2} M*$ there would be observable nuclear burning in the white dwarf phase.  The limits depend on the mass of the white dwarf.
As shown in Table~\ref{tabspectra}, most white dwarfs have atmospheres dominated by H, and the remainder by He. All other elements are only small traces, much less abundant than in the Sun, due to separation in the strong gravitational field\cite{Schatzman48}.
The lightest elements float to the surface once the white dwarf cools below
effective temperature $T_\mathrm{eff} \simeq 100\,000$~K.
Except for the H or He composition, the atmosphere of the white dwarfs in their later cooling evolution has thus lost all memory of the previous evolutionary phases.

We selected candidates to white dwarf stars from the 4.5 million SDSS spectra through their colors, automated spectra fitting by the SDSS pipeline, and targeted search of template spectra.
After complete, consistent human identifications of each candidate white dwarf spectrum, we fitted the optical spectra to local thermodynamic equilibrium (LTE) grids of synthetic non-magnetic 
spectra derived from model atmospheres\cite{Koester10}. Our more recent
DA model grid uses the ML2$/\alpha=0.8$ convection approximation, and for the DBs we use the ML2/$\alpha=1.25$ approximation. The DA grid extends up to $T_\mathrm{eff}=100\,000$~K,
but NLTE effects are not included. Ref.~\refcite{Napiwotzki97} concluded pure hydrogen atmospheres of DA white dwarfs are well represented by LTE calculations for effective temperatures up to 80\,000~K, but when traces of helium are present, non-local thermodynamic equilibrium (NLTE) effects on the Balmer lines occur, down to effective temperatures of 40\,000~K.
Ref.~\refcite{Napiwotzki97} concluded LTE models should exclude traces of helium for the consistent analysis of hydrogen dominated white dwarf spectra.
We fitted the spectral lines and photometry separately\cite{Koester10}, selecting between the hot and cool solutions using photometry as an indicator.
We include corrections to $T_\mathrm{eff}$ and $\log g$ based on three--dimensional convection calculations from Ref.~\refcite{Tremblay13}.

At the cool end of our sample, $\log g=6.5$ corresponds to a mass around $0.2~M_\odot$, well below the single mass
evolution in the lifetime of the Universe. The He-core white dwarf stars in the mass range $0.2-0.45~M_\odot$,
referred to as low-mass white dwarfs, are usually found in close binaries, often double degenerate systems\cite{Marsh95}, being most likely a product of interacting binary star evolution.
More than 70\% of those studied by Ref.~\refcite{Kilic11} with masses below $0.45~M_\odot$ and all but a few with masses below $0.3~M_\odot$ show velocity variations\cite{Brown13,Gianninas14}.
Ref.~\refcite{Kilic07} suggests single low-mass white dwarfs result from the evolution of old metal-rich stars that truncate
evolution before the helium flash due to severe mass loss. They also conclude all white
dwarfs with masses below $\simeq 0.3~M_\odot$ must be a product of binary
star evolution involving interaction between the components, otherwise
the lifetime of the progenitor on the main sequence would be
larger than the age of the Universe.

\section{Spectral fits}

\begin{figure}[pb]
\centerline{\includegraphics[width=\textwidth]{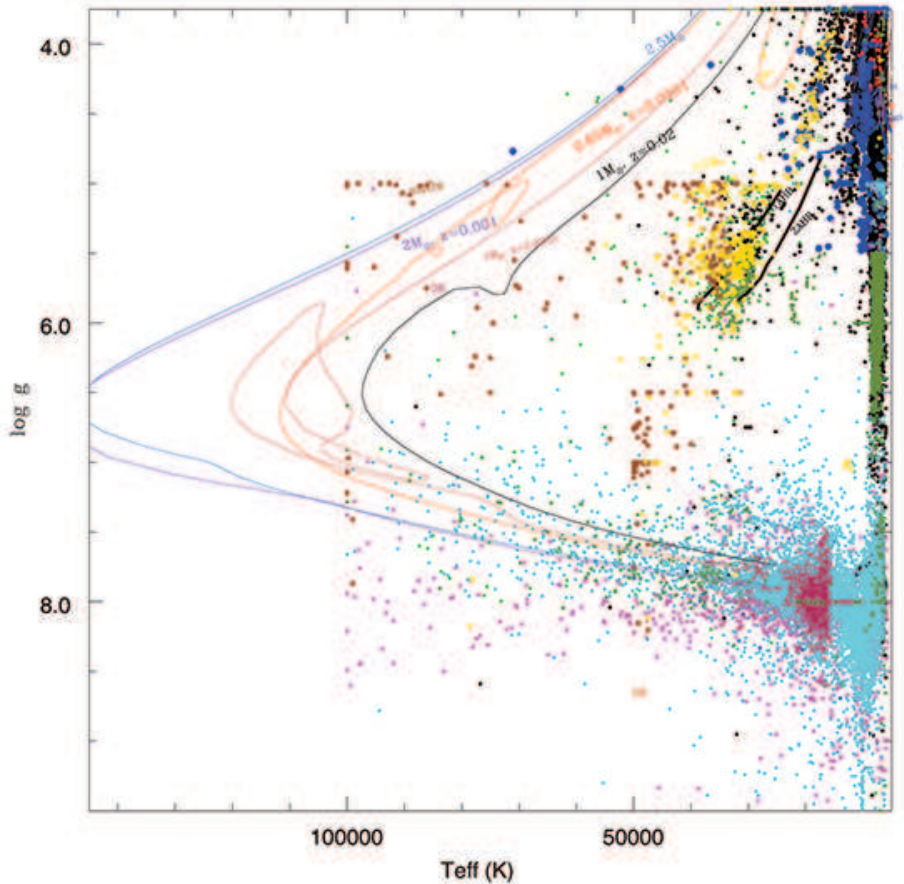}}
\vspace*{8pt}
\caption{Estimated effective temperature and superficial gravity for 79\,440 blue stars in the Sloan
Digital Sky Survey up to Data Release 13, and single star evolutionary models for different metallicities,
showing there is a region in the middle of the diagram that no single star evolutionary models cover.
Data of different colors represent distinct spectral classification and catalogs, DAs in light green, DBs in red, sdBs and sdOs in yellow, sdAs in dark green and black.
The Zero Age Horizontal Branch (ZAHB) and Terminal Age Horizontal Branch (TAHB) plotted were calculated with solar composition models.
\label{single}}
\end{figure}

\begin{figure}[pb]
\centerline{\includegraphics[width=\textwidth]{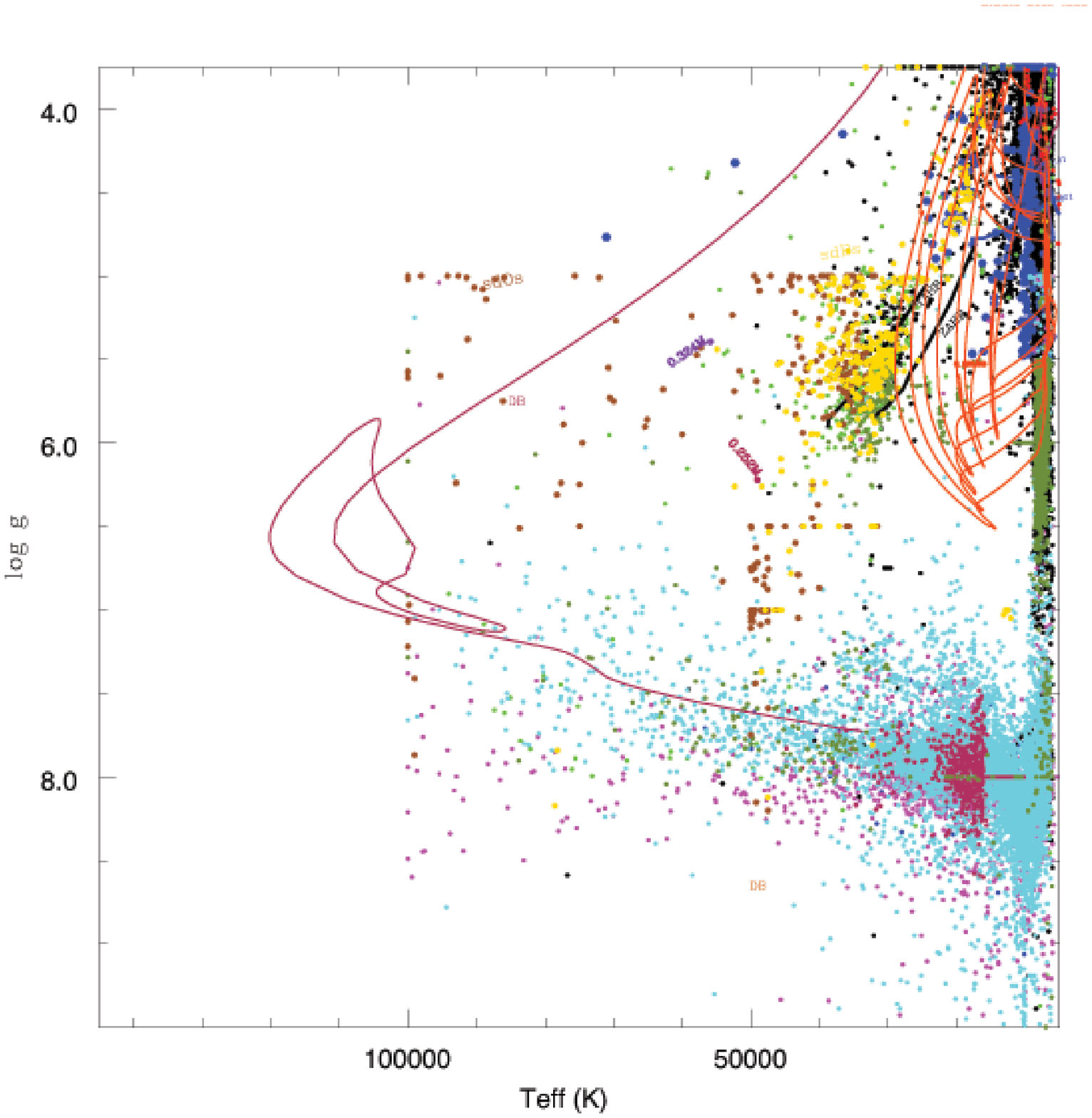}}
\vspace*{8pt}
\caption{Estimated effective temperature and superficial gravity for 79\,440 blue stars in SDSS Data Release 13, and interacting binary star evolutionary models$^{16}$,
showing the region in the middle of the diagram is covered by interacting binary models.
\label{double}}
\end{figure}

In Fig.~\ref{single} we show the results of our effective temperature and surface gravity determinations for all candidates. We calculated the single star evolutionary models shown in the figure
with the MESA\cite{MESA} evolutionary code, including diffusion. In Fig.~\ref{double} the evolutionary models are those with rotation and diffusion of Ref.~\refcite{Istrate16}.

\section{Spectral Classification}

We looked for the following features to aid in the
classification for each specified white dwarf subtype:

\begin{itemize}
\item Balmer lines --- normally broad and with a steep Balmer decrement
[DA but also DAB, DBA, DZA, and subdwarfs]
\item HeI $4\,471$\AA\ [DB, subdwarfs]
\item HeII $4\,686$\AA\ [DO, PG1159, sdO]
\item C2 Swan band or atomic CI lines [DQ]
\item CaII H \& K  [DZ, DAZ, DBZ]
\item CII $4\,367$\AA\ [HotDQ]
\item Zeeman splitting [magnetic white dwarfs]
\item featureless spectrum with significant proper motion [DC]
\item flux increasing in the red [binary, most probably M companion]
\item OI $6\,158$\AA\ [Dox]
\end{itemize}

Table~\ref{tabspectra} shows the spectral classification for the 32\,840 stars we identified as white dwarf stars from SDSS spectra. 84\% are hydrogen atmosphere DAs.
\begin{table}[ph]
\begin{center}
\tbl{Spectral Classification}
{\begin{tabular}{rrl}
Stars&Spectra&Classification\cr 
27661 & 29262 & DAs\cr
2252 & 2403 & DBs\cr
722 & 778 & DAH\cr
1435 & 1494 & DC\cr
935 & 964 & DZ\cr
410 & 440 & DQ\cr
127 & 133 & DO\cr
19 & 21 & DBH\cr
1 & 1 & Dox\cr
\end{tabular} \label{tabspectra}}
\end{center}
\end{table}

\subsection{Dox}
Theoretically, stars with initial masses $7\,M_\odot \leq M \leq 10.6\,M_\odot$
will reach sufficiently high core temperatures,
$T=8$--$12\times 10^8$~K ---
corresponding to center-of-mass energies from 1 to 3~MeV ---
to proceed to carbon burning, and produce either oxygen-neon (ONe) core white
dwarfs, or undergo a core-collapse supernova (SNII) via electron
capture on the products of carbon burning\cite{GarciaBerro94,Doherty14,Nomoto14,Woosley15}. 
The exact outcome of stellar
evolution in this mass range depends critically on the detailed
understanding of the nuclear reaction rates involved, mass--loss, on
the efficiency of convective mixing in the stellar cores\cite{Straniero03}, and metallicity\cite{Romero15}.
Ref.~\refcite{Kepler16} found a white dwarf with oxygen dominated atmosphere,
with no traces of hydrogen, helium or carbon, and with a mass of only
$0.56\pm 0.09~M_\odot$, not expected from evolutionary models, as
such high-mass main sequence stars should result in massive ($M>1~M_\odot$) white dwarfs\cite{Cummings16}.

\subsection{DZs}
Of the white dwarfs in our sample, 3\% have spectra with metal lines, probably
due to accretion of rocky material surrounding the stars\cite{Graham90}\cdash
\cite{Farihi16}.
Calcium and magnesium in general have the strongest lines for white dwarfs at these temperatures.
DZs were the first objects to show asteroidal material exists around stars and Ref.~\refcite{VanderBurg15,Gaensicke16} show
multiple bodies are falling on one white dwarf.

\subsection{Extremely Low Mass white dwarfs}
DA white dwarf stars with masses $M\leq 0.45~M_\odot$
and $T_\mathrm{eff} < 20\,000$~K are Low Mass and Extremely Low Mass (ELM) as found by Refs.
\refcite{Brown10}, \refcite{Kilic11}, \refcite{Brown12}, \refcite{Brown13}, \refcite{Gianninas14}, \refcite{Gianninas15} and \refcite{Brown16}.
Refs.~\refcite{Hermes12} --
\refcite{Bell16} found pulsations
in eight of these ELMs,
similar to the pulsations seen in
DAVs (ZZ~Ceti stars), as described in Ref.~\refcite{VanGrootel13}.
Ref.~\refcite{Maxted14} found 17 pre-ELMs,
i.e., helium--core white dwarf precursors,
and
Ref.~\refcite{Maxted14a,Gianninas16} report pulsations in six of them.
Pulsations are an important tool to study the stellar interior, and
Refs.~\refcite{Corsico14} --
\refcite{Istrate16a} report on theoretical models and pulsations of ELMs.
Refs.~\refcite{Kepler16a} and \refcite{Kepler16b}  show there are thousands of stars, photometrically classified as blue horizontal branch stars by Ref.~\refcite{Xue08,Xue11,Carollo16}, that have spectroscopic
estimated surface gravities much higher than main sequence stars ($\log g \geq 4.75$) and therefore must have radii smaller than the Sun, classifying them as sdAs,
in line with the hot subdwarfs reviewed by Ref.~\refcite{Heber16}.
Ref.~\refcite{Pelisoli16}
discuss they are possibly Extremely Low Mass white dwarf stars.
Refs.~\refcite{Kepler16a,Fusillo15} show that photometrically selected white dwarfs have a contamination around 40\%. Even the ones selected also from proper motion by Ref.~\refcite{Munn17}
show significant contamination by non-white dwarf objects, when spectra are available.

Most stars that produced white dwarfs are born in binaries or multiple systems. Ref.~\refcite{Lada06} demonstrates that while around 70\% of stars more massive than the Sun
are in binaries, two-thirds of the most common stars, M type dwarf stars, are single.
More than 10\% of the spectroscopically identified white dwarfs in SDSS have red companions\cite{Kepler16a,Rebassa16}.
Refs.~\refcite{Farihi10,Nebot11} show that nearly 25\% of all main sequence binaries are close enough that mass transfer interactions occur when the more massive star becomes a red
giant or an asymptotic giant star.
If mass transfer exceeds the Eddington limit, the secondary star is not able to accrete the transferred material and the system evolves through a common envelope
phase, i.e., the core of the giant and the main sequence companion orbit within the
outer layers of the giant star, leading 
to the shrinkage of the orbit and the release of orbital energy. The orbital
energy deposited into the envelope eventually ejects it. Therefore a close binary is formed by the core
of the giant star and a main sequence companion, later a close white dwarf-main sequence binary. 
An ELM will be formed if the envelope is ejected before the helium-flash, which would happen if the star has initial mass too low, i.e., $M\lesssim 2 M_\odot$, 
to reach conditions to fuse helium in the core before it becomes degenerate.

Systems in which a white dwarf is receiving mass from a nearby low mass star are classified as Cataclysmic Variables\cite{Knigge11}.
Accretion of mass by a white dwarf, or the merger of two white dwarfs, may lead to a supernova (SNIa) explosion, if the system grows over the Chandrasekhar mass limit.
The binary white dwarf system SDSS J065133.338+284423.37, with a 12.75~m orbital period, has been identified by Ref.~\refcite{Hermes12a},
showing clear rapid orbital decay consistent with general relativity prediction due to gravitational wave radiation.
White dwarfs have been identified as companions to millisecond pulsars\cite{Durant12}\cdash 
\cite{Bassa16}.

\begin{figure}[pb]
\centerline{\includegraphics[width=0.72\textwidth]{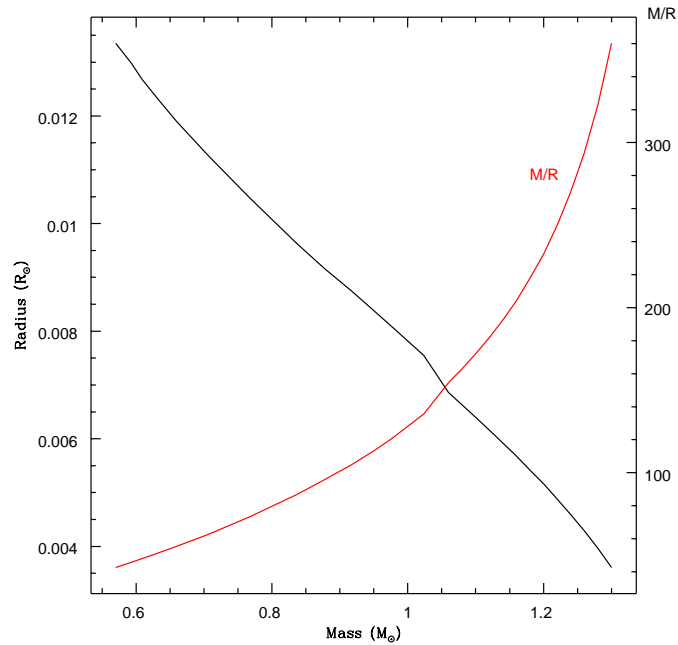}}
\vspace*{8pt}
\caption{Mass and radius or white dwarfs, calculated from C/O e O/Ne evolutionary models at $T_\mathrm{eff}=10\,000$~K.\label{mr}}
\end{figure}

Ref.~\refcite{Tremblay17} uses the Gaia parallaxes for 6 directly observed and 46 white dwarf stars in binary system to estimate their mass-radius relation. By Gaia second data release, in September 2017,
the distance to thousand of white dwarf stars should be known, allowing a precise determination of the mass-radius relation for white dwarfs and clarifying the nature of the Extremely Low Mass white dwarfs and subdwarfs.
\section{Mass Distribution}
We estimated the masses of all DA white dwarfs found by Ref.~\refcite{Kleinman13}, \refcite{Kepler15} and \refcite{Kepler16a}. There were no new optical stellar spectra in SDSS Data Release 13. For the DA mass
distribution, we only consider spectra with S/N$\geq 15$ to have reliable mass determinations. 
We use the
mass--radius relations of Refs.~\refcite{Althaus05}, \refcite{Renedo10} and \refcite{Romero15}
to calculate the mass of our stars from the
$T_\mathrm{eff}$ and  $\log g$ values obtained from our fits,
after correcting to 3D convection following Ref.~\refcite{Tremblay13a}.

\begin{figure}[pb]
\centerline{\includegraphics[width=0.72\textwidth]{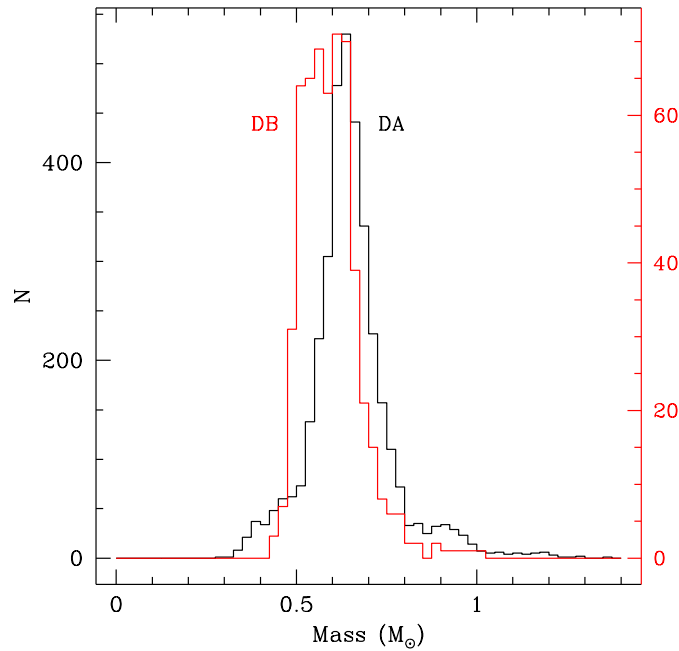}}
\vspace*{8pt}
\caption{Mass distribution by number for 
3636 DAs with $T_\mathrm{eff}\geq 13000$~K, S/N$_g \geq 15$ and $\langle \mathrm{S/N} \rangle=31$ in black and
549 DBs with  $T_\mathrm{eff}\geq 16000$~K, S/N$_g \geq 10$ and  $\langle \mathrm{S/N} \rangle=21$ in red.  \label{dadb}}
\end{figure}

\begin{figure}[pb]
\centerline{\includegraphics[width=0.72\textwidth]{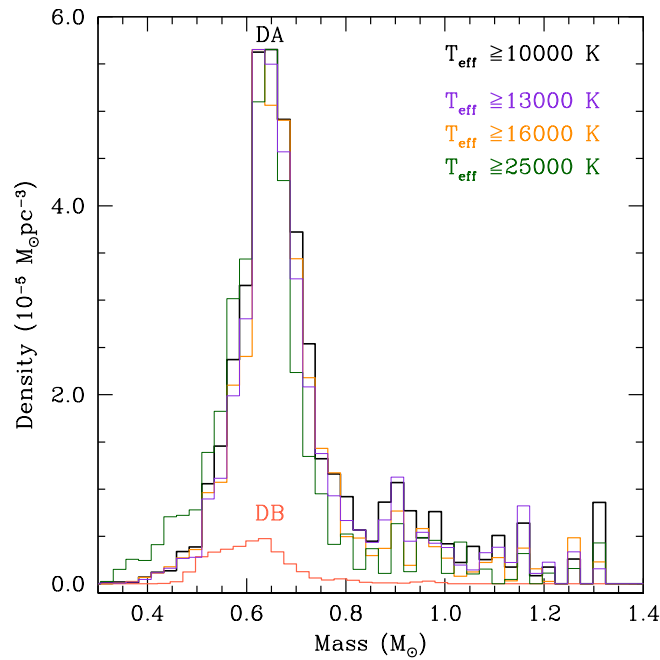}}
\vspace*{8pt}
\caption{Mass distribution corrected by the $1/V_\mathrm{max}$ method for DAs for different cutoff temperatures, and DB with $T_\mathrm{eff} \geq 16\,000$~K.
DAs with $T_\mathrm{eff} \geq 10000$~K, N=4054, $\langle M \rangle=0.647\pm 0.002~M_\odot$ in black,
$T_\mathrm{eff} \geq 13000$~K, N=3637, $\langle M \rangle=0.646\pm 0.002~M_\odot$ in violet,
$T_\mathrm{eff} \geq 16000$~K, N=3012, $\langle M \rangle=0.641\pm 0.002~M_\odot$ in gold,
$T_\mathrm{eff} \geq 25000$~K, N=1121, $\langle M \rangle=0.613\pm 0.003~M_\odot$ in green.
\label{histv10m}}
\end{figure}

Figure~\ref{dadb} shows the mass distribution by number for DAs with $T_\mathrm{eff}\geq 13\,000$~K, where convection is unimportant,
and for DBs with $T_\mathrm{eff}\geq 16\,000$~K reported by Ref.~\refcite{Koester15}.
Because our surface gravities show an unexplained decrease below $T_\mathrm{eff}=10\,000$~K, Figure~\ref{histv10m} shows the 
mass distribution for different cutoff temperatures.

Considering white dwarfs with larger mass have smaller radius, and therefore can only be seen to smaller distances in a magnitude limited survey as SDSS, 
we calculated the density by correcting the visible volume with the $1/V_\mathrm{max}$ method of Ref.~\refcite{Schmidt68},
up to a maximum g=19 magnitude, shown in Figure~\ref{histv10m}.
The distribution shows that the DA and DB distributions have very different shapes. The DA's has a tail to larger masses, while the DB's is extended to lower masses. 
This is probably reflecting some limitation in the progenitors that can undergo very-late thermal pulses and become DBs.

\section{Magnetic Fields}
Ref.~\refcite{GarciaBerro16a} presents a review on magnetic fields in white dwarf stars.
When examining each white dwarf candidate SDSS spectrum by eye, we
found 822 stars with Zeeman splittings indicating magnetic fields
above 2~MG --- the limit where the line splitting becomes too small
to be identified at the SDSS spectral resolution\cite{Kepler13}.

\begin{figure}
\centering
\includegraphics[width=0.8\textwidth]{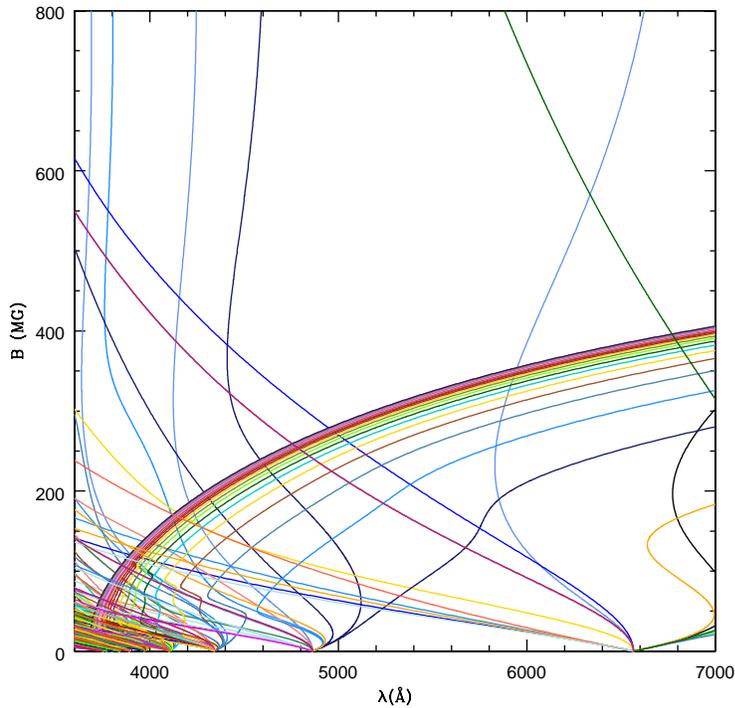}
\caption{Theoretical Zeeman splittings for hydrogen lines in a dipole magnetic field,
showing that even fields around 1~MG produce large splittings of the
higher Balmer lines$^{109-111}$.
\label{chris}}
\end{figure}
If the line splitting and magnetic fields were not recognized, the spectral fittings of DA and DB models would have rendered
too high $\log g$ determinations due to magnetic broadening being misinterpreted as pressure broadening.
We estimated the mean fields for the new DAHs following Ref.~\refcite{Kulebi09} as being from 2~MG to 700~MG.
We caution that stars with large fields are difficult to identify because
fields above around 30~MG
intermixes subcomponents between different hydrogen series components so much that,
depending on effective temperature and signal-to-noise,
it becomes difficult to identify the star as containing hydrogen at all,
and affecting even the colors significantly.
Additionally, white dwarf stars with fields above 100~MG (see Fig.~\ref{chris})
represent the intermediate regime in which the spectra have very few features,
except for a few stationary transitions that have similar wavelengths for a
reasonable distribution of magnetic fields over the surface of the star.
We found significant Zeeman splittings in the spectra of around 4\% of all white dwarfs. 
Both the low field limit and the high
field limit are totally dominated by systematic effects, not the real limits. Polarization is hard to detect because it requires signal-to-noise well above 100, as the fraction of polarized light is
in general smaller than 1\%. Polarimetric studies by Ref.~\refcite{Bagnulo15}
--\refcite{Zejmo17} have shown fields larger than dozens of kilogauss for more than 10\% of the white dwarfs. 
The most massive white dwarf stars have progenitors between O and B main sequence stars. Only one dozen of O-type stars have spectroscopically confirmed magnetic fields, from 50 to 400~G,
while A and B stars have fields from 10 to $10^4$~Gauss\cite{Kholtygin17}.
Even though convection do amplify the magnetic field, most fields are generated during the protostar phase.

The magnetic field detected changes as the star rotates, as in all bodies the rotation axis is not aligned with the magnetic field axis.
SDSS J030407.40-002541.74, 
with g=17.75,
was first classified as a magnetic white dwarf by Ref.~\refcite{Gaensicke02},  
who estimated the magnetic field from one SDSS spectrum, with
Plate-MJD-Fiber=411-51817-172.
They noticed the H$\alpha$ Zeeman triplet split by 200~\AA,
and detected multiple Zeeman components of H$\beta$ and H$\gamma$.
They fitted the spectrum with a 
$T_\mathrm{eff} = 15\,000$ K white
dwarf model, with a dipole field strength $B_d = 10.8$~MG and a magnetic inclination $i = 50^o$
(where $i = 90^o$
corresponds to an
equator-on view). The model spectrum provided a good match
for both the continuum slope and the strength of the Zeeman
absorption lines. The most noticeable shortcoming of their fit was
the poor agreement in the 
$\sigma=-1$
component of H$\beta$. 
Ref.~\refcite{schmidt03} estimated 
$T_\mathrm{eff} = 11\,500$ K 
with a dipole field strength $B_d = 11$~MG and a magnetic inclination $i = 60^o$
from the same spectrum.
Ref.~\refcite{Kulebi09} reanalyzed the same spectrum and fitted an offset dipole with 
$T_\mathrm{eff}=11\,500$~K, assuming $\log g=8.00$, and obtained
$B=11.13 \pm 0.97$~MG, $z_\mathrm{off}=0.27 \pm 0.10\, r_\mathrm{WD}$, and an inclination angle
of $i=52 \pm 17\deg$.
Ref.~\refcite{Kepler13} estimated a varying magnetic field for the 6 SDSS spectra of the star (Table~\ref{sdss}),
assuming dipole magnetic distribution, obtaining B$\simeq$10--19~MG
from the H$\alpha$ and H$\beta$ splittings.

\begin{table}[ph]
\tbl{SDSS spectra and individual spectrum magnetic field estimates.}
{\begin{tabular}{lcrcrcrrrr}
Date Obs.&P-M-F&$B_{H\alpha}$&$B_{H\beta}$&(S/N)$_g$\cr
&&(MG)&(MG)&\cr
2000-09-30 10:36:49 3 exp. 2700s                 &0411-51817-172&11&10&23\cr
2000-11-25 05:47:22 3 exp. 2700s                 &0411-51873-172&19&18&22\cr
2001-01-03 04:33:00 6 exp. 8151s                 &0411-51914-169&10&10&15\cr
2001-10-21 10:24:01 3 exp. 3504s                 &0710-52203-311&11&11&21\cr
2001-10-23 09:46:21 4 exp. 5408s                 &0709-52205-120&11&11&27\cr
2005-01-08 02:55:09 5 exp. 6300s                 &2048-53378-280&10&10&26\cr
\end{tabular}\label{sdss}}
\end{table}

\begin{figure}
\centering
\includegraphics[height=10cm]{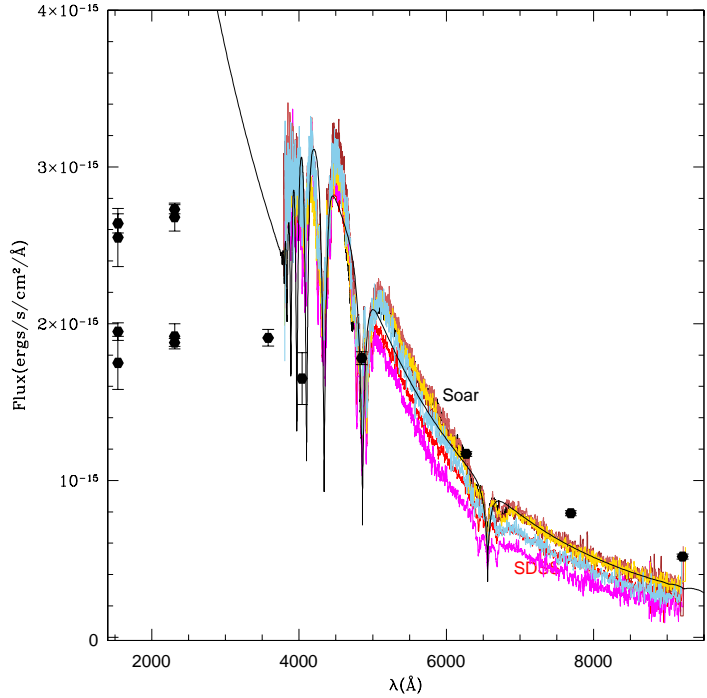}
\caption{Sloan Digital Sky Survey and SOAR spectra of the DAH star
SDSS J030407.40-002541.92, showing the Zeeman splittings changes from spectrum to spectrum.
The thin black line is a non-magnetic model with $T_\mathrm{eff}=11500$~K and $\log g=8$, just to guide the eye.
The photometric measurements are plotted at their central wavelengths. Due to their broad filters,
they represent integrations over absorption lines. The depression in the GALEX colors FUV and NUV
are not caused by absorption lines and are due to the magnetic field. Changes in magnitude are
also seen in the optical.
\label{spectra}}
\end{figure}
A fit to the optical 
SDSS u,g,r colors with non-magnetic models results in
$T_\mathrm{eff}=12\,200 \pm 300$~K, $\log g=8.2\pm  0.1$, and mass
$M= 0.72 \pm 0.06 M_\odot$, all unreliable because the large Zeeman splittings
affect the colors, specially u, substantially.
A fit to the r,i,z colors results in
$T_\mathrm{eff}=13\,500 \pm 500$~K, but the $\log g$ went
to the high border of the grid.
GALEX observations nuv=$18.11 \pm 0.04$, fuv=$18.62 \pm 0.07$
are consistent with
$T_\mathrm{eff}\simeq 12\,000$~K, but the spectra does not
fit from the red to the blue (see Fig.˜\ref{spectra}).

\begin{figure}
\centering
\includegraphics[height=10cm]{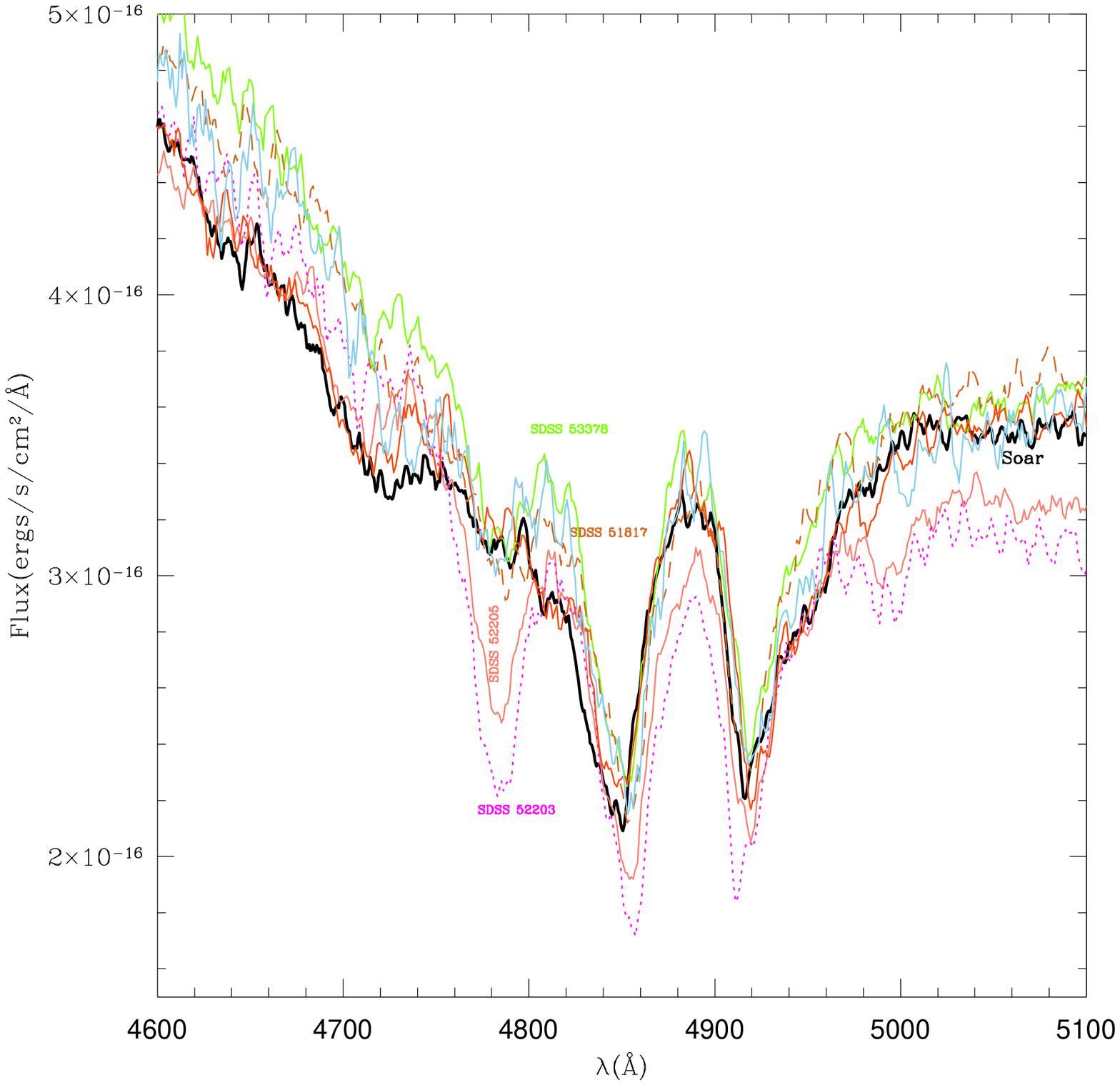}
\caption{Sloan Digital Sky Survey and SOAR spectra of the DAH star
SDSS J030407.40-002541.92 around H$\beta$, showing the Zeeman splitting changes from a triplet to a doublet, at different epochs.
\label{spectrah}}
\end{figure}

The Catalina Real-time Transient
Survey has 311 measurements with a few measurements significantly brighter,
with a mean of $V_{CSS}=17.76\pm 0.05$ and a change from $17.89 \pm 0.11$ to $17.18 \pm 0.16$ in 14 minutes. The timescale of change is of the order of a few hours,
but there is no periodic variation found. The 32 SDSS photometric measurements also show variations in color.

\subsection{Observations}
To test for short timescale variations and obtain a higher signal-to-noise average spectrum, we obtained
$3\times 1800$~s exposures for SDSS J030407.40-002541.92 with the
Soar 4.1~m telescope
on the
night of
2010-09-10 at 06:14:01.77 UT, with a 600~l/mm volume phase holographic grating and a GG-385 red blocking filter on the
Goodman Spectrograph\cite{Clemens04}, reaching a $S/N\simeq 40$ in the continuum. 
Goodman is  mounted  at the  SOAR
Optical  Nasmyth and  its  detector is  a  4k$\times$4k Fairchild  486
back-illuminated   CCD,  with   a   un-binned  plate   scale  of   0.15
arcsec/pixel.
The $\sigma=-1$ component of H$\beta$ was missing,
just like Ref.~\refcite{Gaensicke02} noticed for the Plate-MJD-Fiber=0411-51817-172 SDSS spectra obtained in
2000-09-30.
We did not detect significant changes from spectrum to spectrum (06:14:01, 06:46:06, and 07:17:28 UT).
Ref.~\refcite{Brinkworth13} demonstrated isolated magnetic white dwarfs show photometric variability
on timescale from hours to several years, probably associated with rotation.
Our spectra from SDSS and SOAR is too sparse to estimate the timescale of variability.

\subsection{Analysis}
The field structure
over the surface of the white dwarf becomes accessible to
measurement via the
profiles of the photospheric Zeeman-broadened Balmer absorption
lines, an approach dubbed Zeeman tomography\cite{Euchner02}.
Refs.~\refcite{Euchner05,Euchner06}
show the field geometries of two isolated white dwarfs, 
HE~1045$-$0908 and PG~1015+014,
were 
significantly more complex than simple centered or offset
dipoles.
Ref.~\refcite{Beuermann07} applied the method to cataclysmic variables
and Ref.~\refcite{Kulebi09} applied it to 150 magnetic white dwarf stars observed by SDSS.

The model spectra we fitted  were calculated with a radiative transfer code for magnetized white dwarf atmospheres. For a given temperature and pressure structure of a model atmosphere 
($T_\mathrm{eff}$, $\log g$) and a given magnetic field vector with respect to the line of sight and the normal on the surface of the star,
it calculates theoretical flux and polarization
spectra\cite{Jordan92,Jordan03}.
The model spectra were calculated assuming a surface gravity of $\log g=8$, considering DAs show a sharp peak in the mass distribution around
this value\cite{Kepler07,Gianninas11,Tremblay13}, and we do not have a clear way to estimate the mass of the star. 
Considering no polarization information is available for this star, our analysis is limited to the flux spectra 
(Stokes parameter I). Limb darkening is accounted for by a simple linear scaling law\cite{Euchner02}.
We determine the magnetic field geometry with the code described in Ref.~\refcite{Kulebi09}.
This code calculates the total flux (and circular polarization) spectra for an arbitrary magnetic field topology by adding up appropriately weighted model spectra for a large number of surface elements and then evaluating the goodness of fit. 
Magnetic field geometries are accounted for by multi-pole expansions of the scalar magnetic potential. The individual multi-pole components may be independently oriented with respect to the rotation axis of the white dwarf and offset with respect to its center, allowing in principle for rather complex surface field topologies. 
Additional free parameters are the white dwarf effective temperature and the inclination of the rotation axis with respect to the line of sight. 
The observed spectra was fitted using an evolutionary algorithm\cite{Rechenberg94} with a least-squares quality function.
Additionally to the Zeeman effect, Stark broadening is considered. For the case when the electric and magnetic fields are parallel,
Ref.~\refcite{Friedrich94} estimated the effect on stationary line components, 
which are transitions that vary slowly in wavelength for large intervals of magnetic field strengths. 
Stationary lines are more pronounced than non-stationary lines, as they are not smeared out extensively due to the variation of the magnetic field strength over the stellar surface.
However, no atomic data for hydrogen in the presence of both a magnetic and electric field are available for arbitrary strengths and arbitrary angles between two fields. 
Therefore, only a crude approximation\cite{Jordan92} is used in the model and systematic uncertainties are unavoidable, 
particularly in the low-field regime ($\leq$ 5 MG) where the Stark effect dominates. 
Consequently, effective temperatures and surface gravities derived from fitting the Balmer lines alone are less reliable than in the case of non-magnetic white dwarfs. 
This may also result in disagreements with temperature estimates derived from the continuum slope.
Time-resolved analysis for rotating single magnetic white dwarfs is instrumental in determining rather complex field structures 
[e.g. VLT observations\cite{Euchner02,Euchner05,Euchner06}]. However this usually relies on the preliminary knowledge of period which is usually derived via photometry, 
separately. Although the individual SDSS fiber spectra exists with 15 minute exposure time, 
due to the lack of information on spin period, 
we constrained ourselves to the co-added spectra which includes 3 to 6 individual spectra with total exposure time of at least 45 minutes (see Table~\ref{sdss}).
For white dwarfs, the dipole magnetic field Ohmic decay timescale is $10^{10}$ yr and even the higher multi-poles can live for such a long period of time\cite{Muslimov95}.

\begin{figure}
\centering
\includegraphics[width=0.8\textwidth]{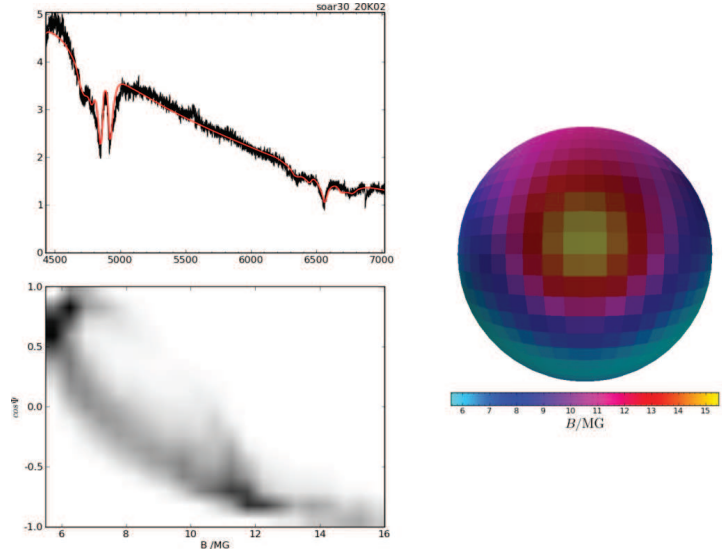}
\caption{Surface magnetic field distribution corresponding to the SOAR average spectrum, calculated by Baybars K\"ulebi.
\label{soar}}
\end{figure}
\begin{figure}
\centering
\includegraphics[width=0.8\textwidth]{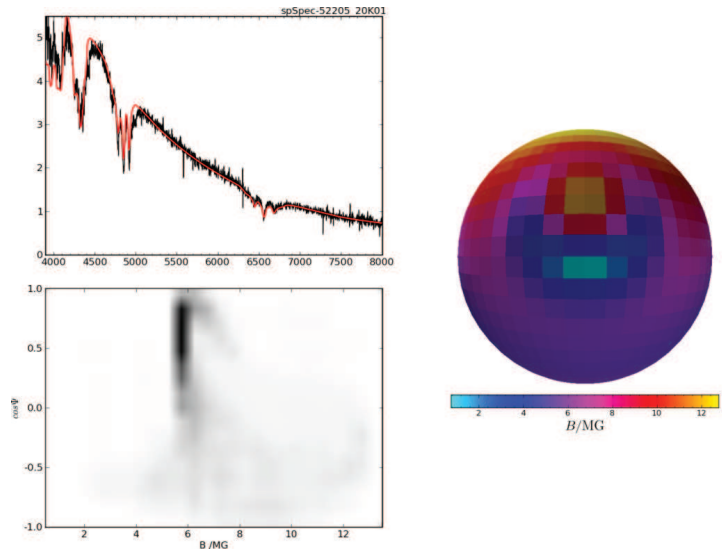}
\caption{Surface magnetic field distribution corresponding to the spectrum shown.
\label{sdsss}}
\end{figure}

Figures~\ref{soar} and \ref{sdsss} shows two graphic forms to present the field distribution across the surface of the star: (i) the
$B \times \cos \phi$ diagram that depict the frequency distribution of field
vectors over the surface of the star corresponding to the spectrum;
and (ii) a synthetic picture of the field distribution.
The latter include (a) the field strength B, (b) $\cos  = B_l/B$ with
$B_l$ the field component along the line of sight.

To extend the calculations of the splittings for
hydrogen levels higher than n=7 for these fields, where perturbation
theory is no longer applicable\cite{Jordan92,Ruder94},
Refs.~\refcite{Schimeczek14,Christopher14}
extended the calculations to n=11.
This was achieved by using a two-dimensional finite element expansion of
the wave functions in terms of $B$-splines in the directions parallel and
perpendicular to the magnetic field, instead of using asymptotically valid
basis expansions in terms of spherical harmonics or Landau orbitals, as
has been previously done in the literature, which fails either in
the high-field or the low-field regime, respectively. In this way
energies could be calculated for 300 states with an accuracy of better than
$10^{-6}$ Rydberg across several symmetry subspaces
over the entire regime of magnetic field strengths. The wealth of the
resulting wavelength information is the basis for the analysis
of the spectra presented in this paper. Details of the computational
method can be found in Ref.~\refcite{Schimeczek13}, and results from 0 to 800~MG are shown in Fig.~\ref{chris}.

The combined effect of electric and
magnetic fields on the spectral lines is very complicated and has been
studied only for special cases of the geometry\cite{Friedrich94,Kulebi09}. Detailed model grids, which
include also the effect of the magnetic field on the radiative
transfer are not yet available.

We demonstrate the Zeeman splitting of the magnetic DA white dwarf
SDSS~J030407.40-002541.92 shows complex field distribution
across the surface of the star, with a mean value of
$B\simeq 13$~MG, and the Zeeman components change
from different spectra, probably due to rotation of the star with
the magnetic axis not aligned to the rotation axis.

Ref.~\refcite{Valyavin14} discuss how magnetic field can suppress convection in white dwarfs, resulting in magnetically generated spots on the stellar surface that can cause the
observed variability, both spectral and photometric, when the star rotates. They also discuss that a magnetic field of the order of the observed in SDSS J030407.40--02541.74
can totally suppress convection, affecting the cooling of the white dwarf. The fast and irregular variations we see in SDSS J030407.40--02541.74 are different from
the stable ones they report for the weaker magnetic field WD~1953--011.

\section{Rotation}
In general the measured rotation period for single white dwarfs ranges from 1~h to 18~d, with a median around 1~d\cite{Kawaler15}.
The fastest single white dwarf rotator from asteroseismological measurements (Table~\ref{rot}) is
the $0.79~M_\odot$ DAV SDSS~J161218.08+083028.1 discovered by
Ref.~\refcite{Castanheira13}, assuming the two observed periods at 115.0~s and 117.0~s are two components of a rotation triplet.

\begin{table}[!ht]
\tbl{Rotation periods of white dwarfs as determined via asteroseismology.}
{\begin{tabular}{lcccc}
\noalign{\smallskip}
Star & $P_{\rm rot}$ [h]  & $T_\mathrm{eff}$ & Type & $M$ [$M_{\odot}$] \\
\noalign{\smallskip}
RX J2117.1+3412  & 28  & 170000 & GW Vir & 0.72 \\
PG 1159-035   & 33  & 140000 & GW Vir & 0.54 \\
NGC 1501  & 28  & 134000 & [WCE] & 0.56 \\
PG 2131+066   &  5  &95000 & GW Vir & 0.55 \\
PG 1707+427   & 16  & 85000 & GW Vir & 0.53 \\
PG 0122+200   & 37  & 80000 & GW Vir & 0.53 \\

PG 0112+104 & 10.17 & 31040 & DBV & 0.58 \\
KIC 8626021 & 43  & 29700 & DBV & 0.56 \\
EC 20058-5234 &  2  & 25500 & DBV & 0.65 \\
GD 358    &   29  & 23740 & DBV & 0.54 \\

G 226-29                 &  9 & 12510 & ZZ Ceti & 0.83 \\
G 185-32                 & 15 & 12470 & ZZ Ceti & 0.67 \\
SDSS J113655.17+040952.6 & 2.6 & 12330 & ZZ Ceti & 0.55 \\
SDSS J161218.08+083028.1 & 0.93 & 12330 & ZZ Ceti & 0.79 \\
Ross 548                 & 37 & 12300 & ZZ Ceti & 0.63 \\
GD 165                   & 50 & 12220 & ZZ Ceti & 0.68 \\
LP 133-144               & 41.8 & 12150 &ZZ Ceti & 0.59 \\
KIC 11911480             & 86.4 & 12160 &ZZ Ceti & 0.58 \\
L 19-2                   & 13 & 12070 & ZZ Ceti & 0.69 \\
HS 0507+0435             & 41   &  12010 & ZZ Ceti & 0.73 \\
EC 14012-1446            & 14.4 & 12020 & ZZ Ceti & 0.72 \\  
KUV 11370+4222           & 5.56 &  11940 & ZZ Ceti & 0.72 \\
G 29-38                  & 32 &   11910 & ZZ Ceti & 0.72 \\
HL Tau 76                & 53   &  11470 & ZZ Ceti & 0.55 \\
SDSS J171113.01+654158.3 & 16.4 &  11130 & ZZ Ceti & 0.90 \\
GD 154                   & 50.4 &  11120 & ZZ Ceti & 0.65 \\
KIC 4552982              & 15.0 &  10860 & ZZ Ceti & 0.71 \\
SDSS J094000.27+005207.1 & 11.8 & 10590 & ZZ Ceti & 0.82 \\
\end{tabular}\label{rot}}
The effective temperatures and masses are corrected to 3D convection\cite{Tremblay13}.
\end{table}

There are two known binary systems with rotation periods around dozens of seconds, called
white dwarf pulsars, 
AE Aquarii and AR Scorpii\cite{Marsh16,Beskrovnaya16}, 
in line with the models of Ref.~\refcite{Malheiro12}.

For magnetic white dwarfs, the $P_\mathrm{rot}=725$~s fast rotator $1.34~M_\odot$ 450~MG DAH EUVE~J0317-855 (1RXS~J031713.9-853231)
listed in Table~2 of Ref.~\refcite{Kawaler15} is in a double degenerate system with the $T_\mathrm{eff}=15600$~K white dwarf LB~9802\cite{Kawka07}. 
Ref.~\refcite{Kawka16} analyzed a non-interacting double white dwarf system NLTT~12758 of a fast spinning ($P_\mathrm{rot}\simeq 23$~m) 3.1~MG 0.69~$M_\odot$ white dwarf and a non-magnetic 0.83~$M_\odot$ white dwarf in a 1.154~d orbit.

Differential rotation in white dwarfs was studied by
Refs.  \refcite{Charpinet09} -- 
\refcite{Hermes16}, 
using the change in rotation splitting of non-radial pulsations, agreeing with the evidence of substantial
loss of angular momentum on the red giant branch from asteroseismology of red giant stars\cite{Cantiello14,Fuller14}.

\section{Axions and Dark Mass}
Axions are the best candidates for dark mass\cite{Ringwald16}. Refs.\refcite{Isern03} --
\refcite{Battich16} show white dwarf pulsations and luminosity function are
consistent with
extra cooling caused by axions of masses around $17\pm 4$~mev.

\section{Population synthesis}
\begin{figure}[pb]
\centerline{\includegraphics[width=0.72\textwidth]{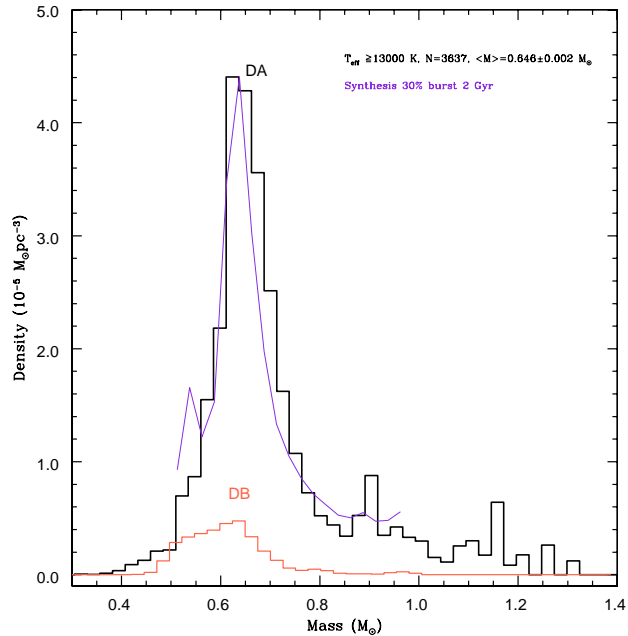}}
\vspace*{8pt}
\caption{Mass distribution corrected by the $1/V_\mathrm{max}$ method for the
3637 DAs with $T_\mathrm{eff} \geq 13\,000$~K,
$\langle M \rangle=0.646\pm 0.002~M_\odot$,
and DBs with $T_\mathrm{eff} \geq 16\,000$~K.
The blue line shows a population synthesis with a 30\% burst 2~Gyr ago, to account for the high mean mass.
The theoretical mass distribution represented by the population synthesis does not include either He-core or O-Ne-core models.
\label{histv13ps}}
\end{figure}

We did a population synthesis analysis and computed a theoretical mass distribution through a Monte Carlo simulation ---
fitting single star initial mass functions, initial-to-final mass relations for masses $0.45~M_\odot \leq  M  < 1.0~M_\odot$ 
--- to obtain a history of star formation
for the DAs. With our calculated lifetimes from ZAMS to the present cooling stage, 
all white dwarfs born 12~Gyr ago are below $T_\mathrm{eff}\simeq 10\,000$˜K,
most are expected to be around 4500~K. For our sample with $T_\mathrm{eff} \geq 13\,000$~K,
Figure~\ref{histv13ps} shows the mean mass around 0.64~$M_\odot$ requires a burst of star formation in the last 2~Gyr,
as a white dwarf with such mass has a short lived progenitor mass with a mass around $2.5~M_\odot$.
This is in contrast with the uniform star formation estimated by Ref.~\refcite{Catalan08} from the ML2/$\alpha=0.6$ mass distribution of Ref.~\refcite{Kepler07}.
These results indicate most white dwarfs are cooler than our present sample and we must
develop ways to disentangle them from red dwarfs. Such cool helium atmosphere white dwarfs show no lines in the spectra, while 
hydrogen atmosphere ones show only a weak H$\alpha$ line. Proper motions and parallaxes are therefore the key separators,
as cool white dwarfs are much fainter than red dwarfs at similar temperatures. Therefore GAIA will be crucial
for their identification.

\section*{Acknowledgments}
SOK, ADR, GO and IP are supported by CNPq-Brazil. This research has made use of NASA's Astrophysics Data System and of the cross-match service provided by CDS, Strasbourg. 
Funding for the Sloan Digital Sky Survey has been provided by the Alfred P. Sloan Foundation, the U.S. Department of Energy Office of Science, and the Participating Institutions. The SDSS web site is www.sdss.org.
Based on observations obtained at the Southern Astrophysical Research (SOAR) telescope, which is a joint project of the Minist\'erio da Ci\^encia, Tecnologia, e Inova\c{c}\~ao (MCTI) da Rep\'ublica Federativa do Brasil, the U.S. National Optical Astronomy Observatory (NOAO), the University of North Carolina at Chapel Hill (UNC), and Michigan State University (MSU).


\section{References}
\begin{thebibliography}{000}
\bibitem{Woosley15} Woosley, S.~E., \& Heger, A.\ 2015, {\it Astrophysical Journal}, 810, 34
\bibitem{Kilic07} Kilic, M., Stanek, K.~Z., \& Pinsonneault, M.~H.\ 2007, {\it Astrophysical Journal}, 671, 761
\bibitem{Romero15} Romero, A.~D., Campos, F., \& Kepler, S.~O.\ 2015, {\it Monthly Notices of the Royal Astronomical Society}, 450, 3708 
\bibitem{GarciaBerro16} Garc{\'{\i}}a-Berro, E., \& Oswalt, T.~D.\ 2016, {\it New Astronomy Reviews}, 72, 1 
\bibitem{Winget87}Winget, D.~E., Hansen, C.~J., Liebert, J., et al.\ 1987, {\it Astrophysical Journal Letters}, 315, L77 
\bibitem{Campos16} Campos, F., Bergeron, P., Romero, A.~D., et al.\ 2016, {\it Monthly Notices of the Royal Astronomical Society}, 456, 3729 
\bibitem{Schatzman48} Schatzman, E.\ 1948,{\it Nature}, 161, 61 
\bibitem{Koester10} Koester, D.\ 2010, {\it Memorie della Societa Astronomica Italiana}, 81, 921
\bibitem{Napiwotzki97} Napiwotzki, R.\ 1997, {\it Astronomy \& Astrophysics}, 322, 256
\bibitem{Tremblay13} Tremblay, P.-E., Ludwig, H.-G., Steffen, M., \& Freytag, B.\ 2013, {\it Astronomy \& Astrophysics}, 552, A13 
\bibitem{Marsh95} Marsh, T.~R., Dhillon, V.~S., \& Duck, S.~R.\ 1995, {\it Monthly Notices of the Royal Astronomical Society}, 275, 828
\bibitem{Kilic11} Kilic, M., Brown, W.~R., Allende Prieto, C., et al.\ 2011, {\it Astrophysical Journal}, 727, 3 
\bibitem{Brown13} Brown W.~R., Kilic M., Allende Prieto C., Gianninas A., Kenyon S.~J., 2013, {\it Astrophysical Journal}, 769, 66
\bibitem{Gianninas14} Gianninas, A., Hermes, J.~J., Brown, W.~R., et al.\ 2014, {\it Astrophysical Journal}, 781, 104 
\bibitem{MESA} Paxton, B., Marchant, P., Schwab, J., et al.\ 2015, {\it Astrophysical Journal Supplement Series}, 220, 15
\bibitem{Istrate16} Istrate, A.~G., Marchant, P., Tauris, T.~M., et al.\ 2016, {\it Astronomy \& Astrophysics}, 595, A35 
\bibitem{GarciaBerro94} Garcia-Berro, E., \& Iben, I.\ 1994, {\it Astrophysical Journal}, 434, 306 
\bibitem{Doherty14} Doherty C.~L., Gil-Pons P., Siess L., Lattanzio J.~C., Lau H.~H.~B., 2015, {\it Monthly Notices of the Royal Astronomical Society}, 446, 2599.
\bibitem{Nomoto14} Nomoto, K.\ 2014, {\it IAU Symposium Supernova Environmental Impacts}, 296, 27
\bibitem{Cummings16} Cummings, J.~D., Kalirai, J.~S., Tremblay, P.-E., Ramirez-Ruiz, E., \& Bergeron, P.\ 2016, {\it Astrophysical Journal Letters}, 820, L18
\bibitem{Straniero03} Straniero, O., Dom{\'{\i}}nguez, I., Imbriani, G., \& Piersanti, L.\ 2003, {\it Astrophysical Journal}, 583, 878 
\bibitem{Kepler16} Kepler, S.~O., Koester, D., \& Ourique, G.\ 2016, {\it Science}, 352, 67 
\bibitem{Graham90} Graham J.~R., Matthews K., Neugebauer G., Soifer B.~T., 1990, {\it Astrophysical Jornal}, 357, 216
\bibitem{Jura03} Jura M., 2003, {\it Astrophysical Journal}, 584, L91
\bibitem{Koester14} Koester D., G{\"a}nsicke B.~T., Farihi J., 2014, {\it Astronomy \& Astrophysics}, 566, A34
\bibitem{Farihi16} Farihi, J.\ 2016, {\it New Astronomy Reviews}, 71, 9
\bibitem{VanderBurg15} Vanderburg, A., Johnson, J.~A., Rappaport, S., et al.\ 2015, {\it Nature}, 526, 546
\bibitem{Gaensicke16} G{\"a}nsicke, B.~T., Aungwerojwit, A., Marsh, T.~R., et al.\ 2016, {\it Astrophysical Journal Letters}, 818, L7 
\bibitem{Brown10} Brown, W.~R., Kilic, M., Allende Prieto, C., \& Kenyon, S.~J.\ 2010, {\it Astrophysical Journal}, 723, 1072 
\bibitem{Brown12} Brown W.~R., Kilic M., Allende Prieto C., Kenyon S.~J., 2012, {\it Astrophysical Journal}, 744, 142
\bibitem{Gianninas15} Gianninas, A., Kilic, M., Brown, W.~R., Canton, P., \& Kenyon, S.~J.\ 2015, {\it Astrophysical Journal}, 812, 167 
\bibitem{Brown16} Brown, W.~R., Gianninas, A., Kilic, M., Kenyon, S.~J., \& Allende Prieto, C.\ 2016, {\it Astrophysical Journal}, 818, 155 
\bibitem{Hermes12} Hermes, J.~J., Montgomery, M.~H., Winget, D.~E., et al.\ 2012, {\it Astrophysical Jornal Letters}, 750, L28 
\bibitem{Hermes13} Hermes, J.~J., Montgomery, M.~H., Winget, D.~E., et al.\ 2013, {\it Astrophysical Jornal}, 765, 102 
\bibitem{Hermes13a} Hermes, J.~J., Montgomery, M.~H., Gianninas, A., et al.\ 2013, {\it Monthly Notices of the Royal Astronomical Society}, 436, 3573 
\bibitem{Bell15} Bell, K.~J., Kepler, S.~O., Montgomery, M.~H., et al.\ 2015, 19th European Workshop on White Dwarfs, 493, 217 
\bibitem{Bell16} Bell, K.~J., Gianninas, A., Hermes, J.~J., et al.\ 2016, arXiv:1612.06390 
\bibitem{VanGrootel13} Van Grootel, V., Fontaine, G., Brassard, P., \& Dupret, M.-A.\ 2013, {\it Astrophysical Journal}, 762, 57
\bibitem{Maxted14} Maxted, P.~F.~L., Bloemen, S., Heber, U., et al.\ 2014, {\it Monthly Notices of the Royal Astronomical Society}, 437, 1681 
\bibitem{Maxted14a} Maxted, P.~F.~L., Serenelli, A.~M., Marsh, T.~R., et al.\ 2014, {\it Monthly Notices of the Royal Astronomical Society}, 444, 208 
\bibitem{Gianninas16} Gianninas, A., Curd, B., Fontaine, G., Brown, W.~R., \& Kilic, M.\ 2016, {\it Astrophysical Journal Letters}, 822, L27 
\bibitem{Corsico14} C{\'o}rsico, A.~H., \& Althaus, L.~G.\ 2014, {\it Astronomy \& Astrophysics}, 569, A106 
\bibitem{Corsico14a} C{\'o}rsico, A.~H., \& Althaus, L.~G.\ 2014, {\it Astrophysical Journal Letters}, 793, L17 
\bibitem{Istrate14} Istrate, A.~G., Tauris, T.~M., \& Langer, N.\ 2014, {\it Astronomy \& Astrophysics}, 571, A45 
\bibitem{Istrate14a} Istrate, A.~G., Tauris, T.~M., Langer, N., \& Antoniadis, J.\ 2014, {\it Astronomy \& Astrophysics}, 571, L3 
\bibitem{Corsico15} C{\'o}rsico, A.~H., \& Althaus, L.~G.\ 2015, 19th European Workshop on White Dwarfs, 493, 221 
\bibitem{Corsico16a} C{\'o}rsico, A.~H., Althaus, L.~G., Calcaferro, L.~M., et al.\ 2016, arXiv:1609.01352 
\bibitem{Istrate16a} Istrate, A.~G., Fontaine, G., Gianninas, A., et al.\ 2016, {\it Astronomy \& Astrophysics}, 595, L12 
\bibitem{Kepler16a} Kepler, S.~O., Pelisoli, I., Koester, D., et al.\ 2016, {\it Monthly Notices of the Royal Astronomical Society}, 455, 3413 
\bibitem{Kepler16b} Kepler, S.~O., Koester, D., Romero, A.~D., Ourique, G., \& Pelisoli, I.\ 2016, arXiv:1610.00371 
\bibitem{Xue08} Xue, X.~X., Rix, H.~W., Zhao, G., et al.\ 2008, {\it Astrophysical Journal}, 684, 1143-1158 
\bibitem{Xue11} Xue, X.-X., Rix, H.-W., Yanny, B., et al.\ 2011, {\it Astrophysical Journal}, 738, 79 
\bibitem{Carollo16} Carollo, D., Beers, T.~C., Placco, V.~M., et al.\ 2016, {\it Nature Physics}, 12, 1170
\bibitem{Heber16} Heber, U.\ 2016, {\it Publications of the Astronomical Society of the Pacific}, 128, 082001 
\bibitem{Pelisoli16} Pelisoli, I., Kepler, S.~O., Koester, D., \& Romero, A.~D.\ 2016, arXiv:1610.05550 
\bibitem{Fusillo15} Gentile Fusillo N.~P., G{\"a}nsicke B.~T., Greiss S., 2015,  {\it Monthly Notices of the Royal Astronomical Society}, 448, 2260
\bibitem{Munn17} Munn, J.~A., Harris, H.~C., von Hippel, T., et al.\ 2017, {\it Astronomical Journal}, 153, 10 
\bibitem{Lada06} Lada, C.~J.\ 2006, {\it Astrophysical Journal Letters}, 640, L63 
\bibitem{Rebassa16} Rebassa-Mansergas, A., Ren, J.~J., Parsons, S.~G., et al.\ 2016, {\it Monthly Notices of the Royal Astronomical Society}, 458, 3808
\bibitem{Farihi10} Farihi, J., Hoard, D.~W., \& Wachter, S.\ 2010, {\it Astrophysical Jornal Supplement Series}, 190, 275
\bibitem{Nebot11} Nebot G{\'o}mez-Mor{\'a}n, A., G{\"a}nsicke, B.~T., Schreiber, M.~R., et al.\ 2011, {\it Astronomy \& Astrophysics}, 536, A43
\bibitem{Knigge11} Knigge, C., Baraffe, I., \& Patterson, J.\ 2011, {\it Astrophysical Jornal Supplement Series}, 194, 28 
\bibitem{Hermes12a}Hermes, J.~J., Kilic, M., Brown, W.~R., et al.\ 2012, {\it Astrophysical Journal Letters}, 757, L21 
\bibitem{Durant12} Durant, M., Kargaltsev, O., Pavlov, G.~G., et al.\ 2012, {\it Astrophysical Journal}, 746, 6 
\bibitem{Tauris12} Tauris, T.~M., Langer, N., \& Kramer, M.\ 2012, {\it Monthly Notices of the Royal Astronomical Society}, 425, 1601 
\bibitem{Antoniadis14} Antoniadis, J.\ 2014, {\it Astrophysical Journal Letters}, 797, L24 
\bibitem{Jia14} Jia, K., \& Li, X.-D.\ 2014, {\it Astrophysical Journal}, 791, 127 
\bibitem{Kaplan14} Kaplan, D.~L., Boyles, J., Dunlap, B.~H., et al.\ 2014, {\it Astrophysical Journal}, 789, 119 
\bibitem{Smedley14} Smedley, S.~L., Tout, C.~A., Ferrario, L., \& Wickramasinghe, D.~T.\ 2014, {\it Monthly Notices of the Royal Astronomical Society}, 437, 2217 
\bibitem{Tauris14} Tauris, T.~M., \& van den Heuvel, E.~P.~J.\ 2014, {\it Astrophysical Journal Letters}, 781, L13 
\bibitem{Cadelano15} Cadelano, M., Pallanca, C., Ferraro, F.~R., et al.\ 2015, {\it Astrophysical Journal}, 812, 63 
\bibitem{Jiang15} Jiang, L., Li, X.-D., Dey, J., \& Dey, M.\ 2015, {\it Astrophysical Journal}, 807, 41 
\bibitem{Kilic15} Kilic, M., Hermes, J.~J., Gianninas, A., \& Brown, W.~R.\ 2015, {\it Monthly Notices of the Royal Astronomical Society}, 446, L26 
\bibitem{Antoniadis16} Antoniadis, J., Kaplan, D.~L., Stovall, K., et al.\ 2016, {\it Astrophysical Journal}, 830, 36 
\bibitem{Deller16} Deller, A.~T., Vigeland, S.~J., Kaplan, D.~L., et al.\ 2016, {\it Astrophysical Journal}, 828, 8 
\bibitem{Bassa16} Bassa, C.~G., Antoniadis, J., Camilo, F., et al.\ 2016, {\it Monthly Notices of the Royal Astronomical Society}, 455, 3806 
\bibitem{Tremblay17} Tremblay, P.-E., Gentile-Fusillo, N., Raddi, R., et al.\ 2017, {\it Monthly Notices of the Royal Astronomical Society}, 465, 2849 
\bibitem{Kleinman13} Kleinman, S.~J., Kepler, S.~O., Koester, D., et al.\ 2013, {\it Astrophysical Journal}s, 204, 5 
\bibitem{Kepler15} Kepler, S.~O., Pelisoli, I., Koester, D., et al.\ 2015, {\it Monthly Notices of the Royal Astronomical Society}, 446, 4078 
\bibitem{Althaus05} Althaus, L.~G., Garc{\'{\i}}a-Berro, E., Isern, J., \& C{\'o}rsico, A.~H.\ 2005, {\it Astronomy \& Astrophysics}, 441, 689 
\bibitem{Renedo10} Renedo, I., Althaus, L.~G., Miller Bertolami, M.~M., et al.\ 2010, {\it Astrophysical Journal}, 717, 183
\bibitem{Tremblay13a} Tremblay P.-E., Ludwig H.-G., Steffen M., Freytag B., 2013, {\it Astronomy \& Astrophysics}, 559, A104 
\bibitem{Koester15} Koester, D., \& Kepler, S.~O.\ 2015, {\it Astronomy \& Astrophysics}, 583, A86 
\bibitem{Schmidt68} Schmidt, M.\ 1968, {\it Astrophysical Journal}, 151, 393
\bibitem{GarciaBerro16a} Garc{\'{\i}}a-Berro, E., Kilic, M., \& Kepler, S.~O.\ 2016, International Journal of Modern Physics D, 25, 1630005 
\bibitem{Kepler13} Kepler S.~O., et al., 2013, {\it Monthly Notices of the Royal Astronomical Society}, 429, 2934 
\bibitem{Kulebi09} K\"ulebi B., Jordan S., Euchner F., G\"ansicke B. T., Hirsch H. 2009, {\it Astronomy \& Astrophysics}, 506, 1341
\bibitem{Bagnulo15a} Bagnulo, S., \& Landstreet, J.~D.\ 2015, {\it Polarimetry of Stars and Planetary Systems}, 224
\bibitem{Bagnulo15} Bagnulo, S., Fossati, L., Landstreet, J.~D., \& Izzo, C.\ 2015, {\it Astronomy \& Astrophysics}, 583, A115 
\bibitem{Zejmo17} {\.Z}ejmo, M., S{\l}owikowska, A., Krzeszowski, K., Reig, P., \& Blinov, D.\ 2017, {\it Monthly Notices of the Royal Astronomical Society}, 464, 1294 
\bibitem{Kholtygin17} Kholtygin, A.~F., Fabrika, S., Hubrig, S., et al.\ 2017, arXiv:1701.00739 
\bibitem{Gaensicke02} G{\"a}nsicke B.~T., Euchner F., Jordan S., 2002, {\it Astronomy \& Astrophysics}, 394, 957 
\bibitem{schmidt03} Schmidt G.~D., et al., 2003, {\it Astrophysical Journal}, 595, 1101 
\bibitem{Clemens04} Clemens J.~C., Crain J.~A., Anderson R., 2004, SPIE, 5492, 331
\bibitem{Brinkworth13} Brinkworth C.~S., Burleigh M.~R., Lawrie K., Marsh T.~R., Knigge C., 2013, {\it Astrophysical Journal}, 773, 47 
\bibitem{Euchner02} Euchner F., Jordan S., Beuermann K., G{\"a}nsicke B.~T., Hessman F.~V., 2002, {\it Astronomy \& Astrophysics}, 390, 633 
\bibitem{Euchner05} Euchner F., Reinsch K., Jordan S., Beuermann K., G{\"a}nsicke B.~T., 2005, {\it Astronomy \& Astrophysics}, 442, 651 
\bibitem{Euchner06} Euchner F., Jordan S., Beuermann K., Reinsch K., G{\"a}nsicke B.~T., 2006, {\it Astronomy \&A Astrophysics}, 451, 671 
\bibitem{Beuermann07} Beuermann K., Euchner F., Reinsch K., Jordan S., G{\"a}nsicke B.~T., 2007, {\it Astronomy \& Astrophysics}, 463, 647 
\bibitem{Jordan92} Jordan S., 1992, {\it Astronomy \& Astrophysics}, 265, 570
\bibitem{Jordan03} Jordan S., Schmidt H., 2003, ASPC, 288, 625
\bibitem{Jordan07} Jordan S., Aznar Cuadrado R., Napiwotzki R., Schmid H. M., Solanki S. K. 2007, {\it Astronomy \& Astrophysics}, 462, 1097
\bibitem{Kepler07} Kepler, S.~O., Kleinman, S.~J., Nitta, A., et al.\ 2007, {\it Monthly Notices of the Royal Astronomical Society}, 375, 1315 
\bibitem{Gianninas11} Gianninas A., Bergeron P., Ruiz M.~T., 2011, {\it Astrophysical Journal}, 743, 138 
\bibitem{Rechenberg94} Rechenberg, I. 1994, Werksatt Bionik und Evolutionstechnik No. 1 (Stuttgart:frommann-holsboog)
\bibitem{Friedrich94} Friedrich S., Ostreicher R., Ruder H., Zeller G., 1994, {\it Astronomy \& Astrophysics}, 282, 179 
\bibitem{Muslimov95} Muslimov A.~G., van Horn H.~M., Wood M.~A., 1995, {\it Astrophysical Journal}, 442, 758
\bibitem{Ruder94} Ruder H., Wunner G., Herold H., Geyer F., 1994, Atoms in Strong Magnetic Fields. Quantum Mechanical Treatment and Applications in Astrophysics and Quantum Chaos, Springer-Verlag, Heidelberg.
\bibitem{Schimeczek14} Schimeczek C., Wunner G., 2014a, {\it Computer Physics Communications}, 185, 614 
\bibitem{Christopher14} Schimeczek C., Wunner G., 2014b, {\it Astrophysical Journals Supplement Series}, 212, 26 
\bibitem{Schimeczek13} Schimeczek C., Boblest S., Meyer D., Wunner G., 2013, {\it Physical Review A}, 88, 012509 
\bibitem{Valyavin14} Valyavin G., et al., 2014, Nature, 515, 88 
\bibitem{Kawaler15} Kawaler, S.~D.\ 2015, 19th European Workshop on White Dwarfs, ASPC 493, 65 
\bibitem{Kawka07} Kawka, A., Vennes, S., Schmidt, G.~D., Wickramasinghe, D.~T., \& Koch, R.\ 2007, {\it Astrophysical Journal}, 654, 499 
\bibitem{Castanheira13} Castanheira, B.~G., Kepler, S.~O., Kleinman, S.~J., Nitta, A., \& Fraga, L.\ 2013, {\it Monthly Notices of the Royal Astronomical Society}, 430, 50 
\bibitem{Marsh16} Marsh, T.~R., G{\"a}nsicke, B.~T., H{\"u}mmerich, S., et al.\ 2016, {\it Nature}, 537, 374 
\bibitem{Beskrovnaya16} Beskrovnaya, N.~G., \& Ikhsanov, N.~R.\ 2016, arXiv:1612.07831 
\bibitem{Malheiro12} Malheiro, M., Rueda, J.~A., \& Ruffini, R.\ 2012, {\it Publication of the Astronomical Society of Japan}, 64, 56 
\bibitem{Kawka16} Kawka, A., Briggs, G.~P., Vennes, S., et al.\ 2016, arXiv:1612.00325 
\bibitem{Charpinet09} Charpinet, S., Fontaine, G., \& Brassard, P.\ 2009, {\it Nature}, 461, 501
\bibitem{Corsico11} C{\'o}rsico, A.~H., Althaus, L.~G., Kawaler, S.~D., et al.\ 2011, {\it Monthly Notices of the Royal Astronomical Society}, 418, 2519
\bibitem{Fontaine13} Fontaine, G., Brassard, P., \& Charpinet, S.\ 2013, 18th European White Dwarf Workshop., 469, 115 
\bibitem{Hermes16} Hermes, J.~J., Kawaler, S.~D., Bischoff-Kim, A., et al.\ 2016, arXiv:1612.07807 
\bibitem{Cantiello14} Cantiello, M., Mankovich, C., Bildsten, L., Christensen-Dalsgaard, J., \& Paxton, B.\ 2014, {\it Astrophysical Journal}, 788, 93 
\bibitem{Fuller14} Fuller, J., Lecoanet, D., Cantiello, M., \& Brown, B.\ 2014, {\it Astrophysical Journal}, 796, 17 
\bibitem{Ringwald16} Ringwald, A.\ 2016, arXiv:1612.08933 
\bibitem{Isern03} Isern, J., \& Garc{\'{\i}}a-Berro, E.\ 2003, {\it Nuclear Physics B Proceedings Supplements}, 114, 107 
\bibitem{Isern10} Isern, J., Garc{\'{\i}}a-Berro, E., Althaus, L.~G., \& C{\'o}rsico, A.~H.\ 2010, {\it Astronomy \& Astrophysics}, 512, A86 
\bibitem{Corsico12a} C{\'o}rsico, A.~H., Althaus, L.~G., Miller Bertolami, M.~M., et al.\ 2012, {\it Monthly Notices of the Royal Astronomical Society}, 424, 2792 
\bibitem{Corsico12} C{\'o}rsico, A.~H., Althaus, L.~G., Romero, A.~D., et al.\ 2012, {\it Journal of Cosmology and Astroparticle Physics}, 12, 010 
\bibitem{Corsico16} C{\'o}rsico, A.~H., Romero, A.~D., Althaus, L.~G., et al.\ 2016, {\it Journal of Cosmology and Astroparticle Physics}, 7, 036 
\bibitem{Battich16} Battich, T., C{\'o}rsico, A.~H., Althaus, L.~G., \& Miller Bertolami, M.~M.\ 2016, {\it Journal of Cosmology and Astroparticle Physics}, 8, 062 
\bibitem{Catalan08} Catal{\'a}n, S., Isern, J., Garc{\'{\i}}a-Berro, E., \& Ribas, I.\ 2008, {\it Monthly Notices of the Royal Astronomical Society}, 387, 1693 
\end{thebibliography}
\end{document}